\documentclass[aps,twocolumn,amsmath,amssymb,superscriptaddress,prl,showpacs]{revtex4}
\usepackage[dvipdfmx]{graphicx}
\usepackage[dvipdfmx]{color}
\usepackage{amsmath}
\usepackage{amssymb}
\usepackage{dcolumn}
\usepackage{bm}
\usepackage{latexsym} 

\begin{document}

\title{Longitudinal Spin Seebeck Effect Free from the Proximity Nernst Effect}

\author{T. Kikkawa}
\affiliation{Institute for Materials Research, Tohoku University, Sendai 980-8577, Japan}

\author{K. Uchida}
\email{kuchida@imr.tohoku.ac.jp}
\affiliation{Institute for Materials Research, Tohoku University, Sendai 980-8577, Japan}
\affiliation{PRESTO, Japan Science and Technology Agency, Saitama 332-0012, Japan}

\author{Y. Shiomi}
\affiliation{WPI Advanced Institute for Materials Research, Tohoku University, Sendai 980-8577, Japan}

\author{Z. Qiu}
\affiliation{WPI Advanced Institute for Materials Research, Tohoku University, Sendai 980-8577, Japan}

\author{D. Hou}
\affiliation{State Key Laboratory of Surface Physics and Department of Physics, Fudan University, Shanghai 200433, China}

\author{D. Tian}
\affiliation{State Key Laboratory of Surface Physics and Department of Physics, Fudan University, Shanghai 200433, China}

\author{H. Nakayama}
\affiliation{Institute for Materials Research, Tohoku University, Sendai 980-8577, Japan}

\author{X.-F. Jin}
\affiliation{State Key Laboratory of Surface Physics and Department of Physics, Fudan University, Shanghai 200433, China}

\author{E. Saitoh}
\affiliation{Institute for Materials Research, Tohoku University, Sendai 980-8577, Japan}
\affiliation{WPI Advanced Institute for Materials Research, Tohoku University, Sendai 980-8577, Japan}
\affiliation{CREST, Japan Science and Technology Agency, Tokyo 102-0076, Japan}
\affiliation{Advanced Science Research Center, Japan Atomic Energy Agency, Tokai 319-1195, Japan}
\date{\today}
\begin{abstract}
This letter provides evidence for intrinsic longitudinal spin Seebeck effects (LSSEs) that are free from the anomalous Nernst effect (ANE) caused by an extrinsic proximity effect. We report the observation of LSSEs in Au/Y$_3$Fe$_5$O$_{12}$ (YIG) and Pt/Cu/YIG systems, showing that LSSE appears even when the mechanism of the proximity ANE is clearly removed. In the conventional Pt/YIG structure, furthermore, we separate the LSSE from the ANE by comparing the voltages in different magnetization and temperature-gradient configurations; the ANE contamination was found to be negligibly small even in the Pt/YIG structure. 
\end{abstract}
\maketitle
%
%
The spin Seebeck effect (SSE) refers to the generation of a spin voltage as a result of a temperature gradient in magnetic materials~\cite{SSE1,SSE2,SSE3,SSE4,SSE5,SSE6,SSE8,SSE10,SSE11,SSE12,SSE13,SSE14}. Here, a spin voltage is a potential for electron spins to drive a nonequilibrium spin current~\cite{spincurrent1,spincurrent2}; when a conductor is attached to a magnet with a finite spin voltage, it induces a spin injection into the conductor. The SSE is of crucial importance in spintronics~\cite{spintronics1,spintronics2,spintronics3} and spin caloritronics~\cite{spincaloritronics1,spincaloritronics2} since it enables the simple and versatile generation of a spin current from heat. \par
The simplest and most straightforward setup of the SSE is the longitudinal configuration~\cite{SSE5,SSE13}, in which a spin current flowing parallel to a temperature gradient is measured via the inverse spin Hall effect (ISHE)~\cite{ISHE1,ISHE2,ISHE3,ISHE6,ISHE_YIGAu,ISHE_nakayama}. The longitudinal SSE (LSSE) device consists of a ferromagnetic or ferrimagnetic insulator (e.g., Y$_3$Fe$_5$O$_{12}$: YIG) covered with a paramagnetic metal (e.g., Pt) film [Fig. 1(a)]. When a temperature gradient $\nabla T$ is applied to the ferromagnetic insulator perpendicular to the ferromagnet/paramagnet interface ($z$ direction) and the magnetization ${\bf M}$ of the ferromagnet is along the $x$ direction, an ISHE-induced voltage is generated in the paramagnet along the $y$ direction according to the relation
\begin{equation}\label{equ:SSE1}
{\bf E}_{\rm ISHE} \propto {\bf J}_{\rm s} \times {\bm \sigma}, 
\end{equation}
where ${\bf E}_{\rm ISHE}$, ${\bf J}_{\rm s}$, and ${\bm \sigma}$ denote the electric field induced by the ISHE, the spatial direction of the thermally generated spin current, and the spin-polarization vector of electrons in the paramagnet, respectively [Fig. 1(a)]. Therefore, by measuring ${\bf E}_{\rm ISHE}$, the LSSE can be detected electrically. Here, the use of a highly resistive insulator, such as YIG, is indispensable to the LSSE experiments to eliminate artifacts caused by electric conduction in the ferromagnetic layer~\cite{SSE13}. \par
\begin{figure}[t]
\begin{center}
\includegraphics[width=5.3cm]{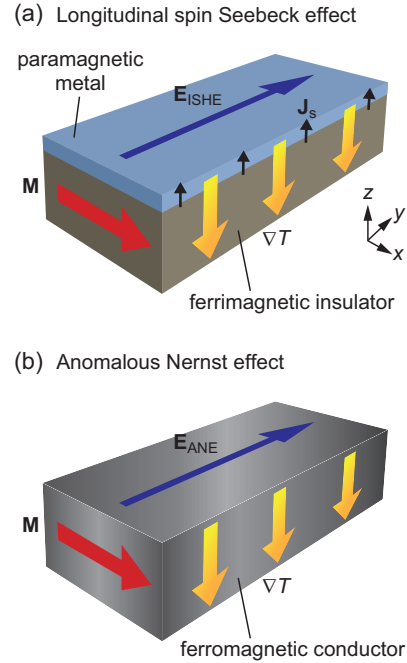}
\caption{Schematic illustrations of (a) the LSSE and (b) the ANE. $\nabla T$, ${\bf M}$, ${\bf E}_{\rm ISHE}$ (${\bf E}_{\rm ANE}$), and ${\bf J}_{\rm s}$ denote the temperature gradient, the magnetization vector, the electric field induced by the ISHE (ANE), and the spatial direction of the thermally generated spin current, respectively. }\label{fig:1}
\end{center}
\end{figure}
The observation of the LSSE has typically been reported in Pt/YIG systems~\cite{SSE5,SSE13}. However, in this conventional structure, since Pt is near the Stoner ferromagnetic instability~\cite{Ibach,DOS}, ferromagnetism may be induced in the Pt layer in the vicinity of the Pt/YIG interface due to a static proximity effect, although the magnetization of YIG is so small that the proximity is expected to be weak. Against this backdrop, Huang {\it et al.} \cite{Chien} pointed out that the LSSE voltage in the Pt/YIG structure might be contaminated by the anomalous Nernst effect (ANE)~\cite{Nernst} due to the magnetic proximity in the Pt layer [compare Figs. \ref{fig:1}(a) and \ref{fig:1}(b)] and that the exclusive establishment of the LSSE is necessary. In this letter, to do that, we provide clear evidence for the existence of the LSSE free from the ANE induced by the proximity effect. First, we show that the LSSE appears even in Au/YIG and Pt/Cu/YIG structures; since Au and Cu are typical metals far from the Stoner instability~\cite{Ibach,DOS}, the observation of the LSSE in these structures allows us to conclude that the LSSE is an intrinsic phenomenon and different from the proximity ANE. Second, we demonstrate that the LSSE can be distinguished from the ANE by comparing the voltages in different magnetization and temperature-gradient configurations and that the ANE contamination is negligibly small even in the Pt/YIG structure. \par
\begin{figure}[t]
\begin{center}
\includegraphics[width=8.5cm]{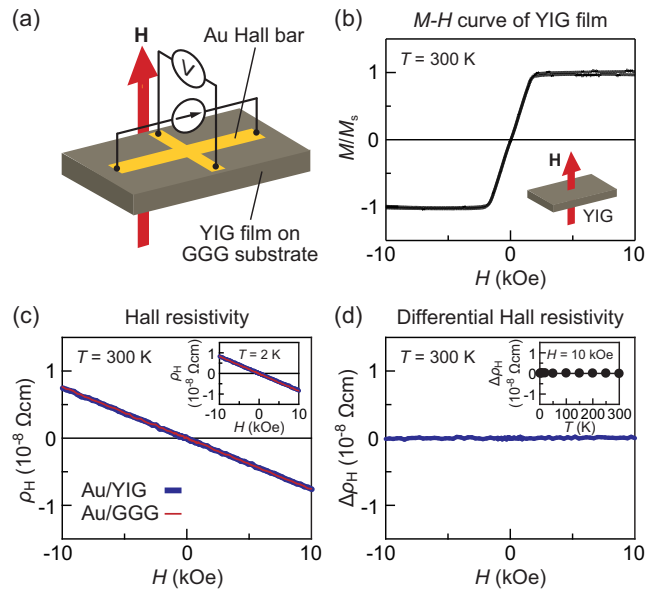}
\caption{(a) A schematic illustration of the Au/YIG sample used for the Hall measurements. The magnetic field ${\bf H}$ with the magnitude $H$ was applied perpendicular to the Au/YIG interface. (b) Magnetization ($M$)-$H$ curve of the YIG film at the temperature $T = 300~\textrm{K}$, measured when ${\bf H}$ was applied perpendicular to the film plane. Here, the vertical axis is normalized by the saturation magnetization $M_{\rm s}$. (c) $H$ dependence of the Hall resistivity $\rho_{\rm H}$ in the Au/YIG and Au/GGG samples at $T = 300~\textrm{K}$. The inset to (c) shows the $H$ dependence of $\rho_{\rm H}$ at $T = 2~\textrm{K}$. (d) $H$ dependence of the differential Hall resistivity $\Delta \rho_{\rm H}$ at $T = 300~\textrm{K}$, where $\Delta \rho_{\rm H}$ denotes the difference of $\rho_{\rm H}$ between the Au/YIG and Au/GGG samples. The inset to (d) shows the $T$ dependence of $\Delta \rho_{\rm H}$ at $H = 10~\textrm{kOe}$. }\label{fig:2}
\end{center}
\end{figure}

First of all, we measured Hall effects in a Au/YIG structure. Here, a 4.5-$\mu$m-thick single-crystalline YIG (111) film was grown on a 0.5-mm-thick Gd$_3$Ga$_5$O$_{12}$ (GGG) (111) substrate by a liquid phase epitaxy method, where the resistance of the YIG film is much greater than the measurement limit of our electrometer: 210 G$\Omega$. A 10-nm-thick Au Hall bar was then deposited on the surface of the YIG film [Fig. \ref{fig:2}(a)]. As shown in Fig. \ref{fig:2}(c), the Hall resistivity in the Au/YIG sample varies linearly with respect to the magnitude of an external magnetic field $H$ applied perpendicular to the Au/YIG interface due to the normal Hall effect in Au. We found that the $H$ dependence of the Hall resistivity in the Au/YIG sample completely coincides with that in a Au/GGG sample, where the Au Hall bar was fabricated directly on a paramagnetic GGG substrate [Figs. \ref{fig:2}(c) and \ref{fig:2}(d)]. The same behavior was observed at temperatures ranging from 300 to 2 K [see the inset to Fig. \ref{fig:2}(d)]. These results confirm that the Au film on YIG exhibits no anomalous Hall effect, a situation enabling the detection of the intrinsic LSSE free from the magnetic proximity. \par
Now we focus on the LSSE in the Au/YIG structure. To measure the LSSE, we used the Au/YIG sample with a 7 $\times$ 6 mm$^2$ rectangular shape, where the whole surface of the YIG film is covered with the Au film. A temperature gradient $\nabla T$ was applied perpendicular to the Au/YIG interface. Here, the temperatures of the top of the Au film and the bottom of the GGG substrate were stabilized, respectively, at $300~\textrm{K}$ and $300~\textrm{K} \pm \Delta T$. During the LSSE measurements, a magnetic field ${\bf H}$ was applied along the $x$ or $y$ direction; the magnetization of the YIG film is aligned along the in-plane direction when $|H| > 30~\textrm{Oe}$. Under this condition, we measured an electric voltage $V$ between the ends of the Au layer along the $y$ direction. \par
\begin{figure}[t]
\begin{center}
\includegraphics[width=8.6cm]{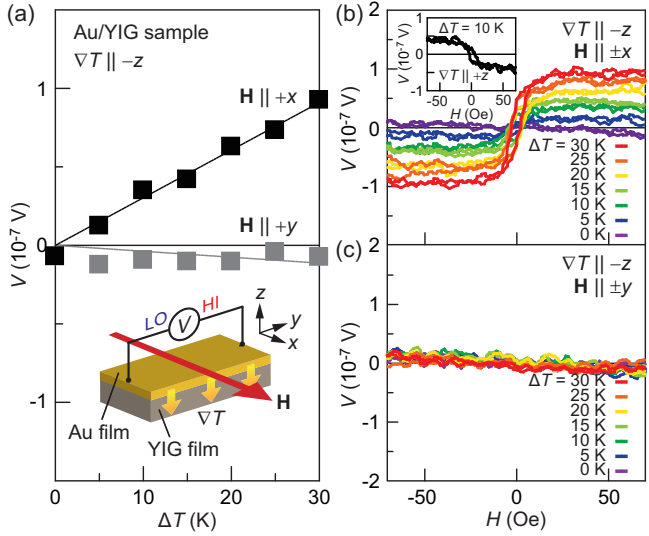}
\caption{(a) Temperature-difference ($\Delta T$) dependence of the voltage $V$ in the Au/YIG sample at $H = 50~\textrm{Oe}$, measured when $\nabla T~|| -z$ and ${\bf H}~|| +x$ or $+y$. The Au/YIG sample was formed on the GGG substrate with the size of $7 \times 6 \times 0.5$ mm$^3$, where the thicknesses of the Au and YIG films are 10 nm and 4.5 $\mu$m, respectively. (b) $H$ dependence of $V$ in the Au/YIG sample for various values of $\Delta T$, measured when $\nabla T~|| -z$ and ${\bf H}~|| \pm x$. The inset to (b) shows the $H$ dependence of $V$ at $\Delta T = 10~\textrm{K}$, measured when $\nabla T~|| +z$ and ${\bf H}~|| \pm x$. (c) $H$ dependence of $V$ in the Au/YIG sample for various values of $\Delta T$, measured when $\nabla T~|| -z$ and ${\bf H}~|| \pm y$. }\label{fig:3}
\end{center}
\end{figure}

Figure \ref{fig:3}(a) shows the measured $V$ in the Au/YIG sample at $H = 50~\textrm{Oe}$ as a function of the temperature difference $\Delta T$. When ${\bf H}$ was applied along the $x$ direction, the $V$ signal was observed to be proportional to $\Delta T$. We confirmed that the sign of the $V$ signal is reversed by reversing $H$ [Fig. \ref{fig:3}(b)] and that the signal disappears when ${\bf H}$ is along the $y$ direction [Figs. \ref{fig:3}(a) and \ref{fig:3}(c)]. By reversing the $\nabla T$ direction, the sign of $V$ is also reversed [see the inset to Fig. \ref{fig:3}(b)]. This behavior of the $V$ signal in the Au/YIG sample is the characteristic of the ISHE voltage induced by the LSSE. Since Au is a typical metal far from ferromagnetism, this result shows that the LSSE appears even in the absence of the proximity ANE. \par
\begin{figure}[t]
\begin{center}
\includegraphics[width=8.6cm]{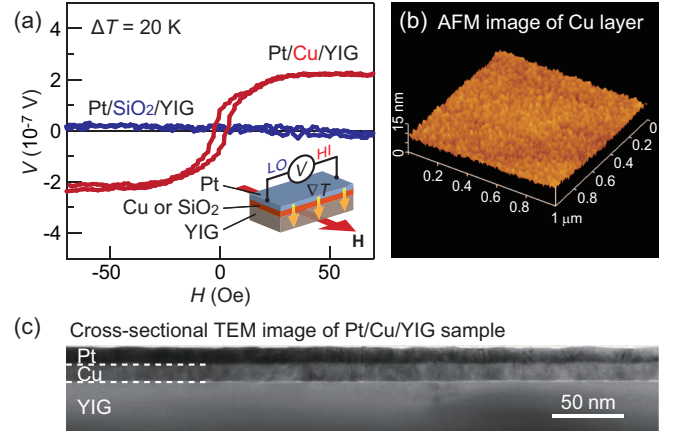}
\caption{(a) $H$ dependence of $V$ in the Pt/Cu/YIG and Pt/SiO$_2$/YIG samples at $\Delta T = 20~\textrm{K}$, measured when $\nabla T~|| -z$ and ${\bf H}~|| \pm x$. The Pt/Cu/YIG and Pt/SiO$_2$/YIG samples were formed on the GGG substrate with the size of $7 \times 6 \times 0.5$ mm$^3$, where the thicknesses of the Pt, Cu, SiO$_2$, and YIG films are 10 nm, 13 nm, 10 nm, and 4.5 $\mu$m, respectively. (b) An atomic-force-microscope (AFM) image of the surface of the Cu film fabricated on the YIG film, where the surface roughness is $R_{\rm a} = 0.67~\textrm{nm}$. (c) A cross-sectional transmission-electron-microscope (TEM) image of the Pt/Cu/YIG sample. }\label{fig:4}
\end{center}
\end{figure}

Next, we show that the LSSE appears also in the Pt/Cu/YIG structure, where a Cu film is inserted between Pt and YIG films to block the possible induced ferromagnetism in Pt. The thickness of the Cu interlayer is 13 nm; this is thick enough to eliminate the magnetic proximity between Pt and YIG, while the LSSE voltage can persist in the Pt layer owing to the long spin-diffusion length of Cu \cite{ISHE3}. We checked that the surface roughness of the Cu interlayer is smaller than 1 nm by means of atomic force microscopy [Fig. \ref{fig:4}(b)] and transmission electron microscopy [Fig. \ref{fig:4}(c)], a result showing that the Pt/Cu/YIG sample has no pinholes in the Cu interlayer and no leaked magnetic proximity between the Pt and YIG layers. As shown in Fig. \ref{fig:4}(a), we also observed a clear $V$ signal in the Pt/Cu/YIG sample, providing further evidence for the existence of the intrinsic LSSE since Cu is also free from induced ferromagnetism. In contrast, we found that the $V$ signal disappears in a Pt/SiO$_2$/YIG sample, where the Cu layer is replaced by an insulating 10-nm-thick SiO$_2$ film [Fig. \ref{fig:4}(a)]. This result confirms that 10 nm is thick enough to eliminate the magnetic proximity between Pt and YIG and that the direct contact between YIG and metal is necessary for generating the spin current, consistent with the scenario of the LSSE. \par
The above experiments in the Au/YIG and Pt/Cu/YIG samples clearly show the existence of the intrinsic LSSE free from the ANE. Then, is the voltage signal in the conventional Pt/YIG structure contaminated by the proximity ANE? The following experiments provide a clear-cut answer to this question. \par
\begin{figure*}[t]
\begin{center}
\includegraphics[width=15.5cm]{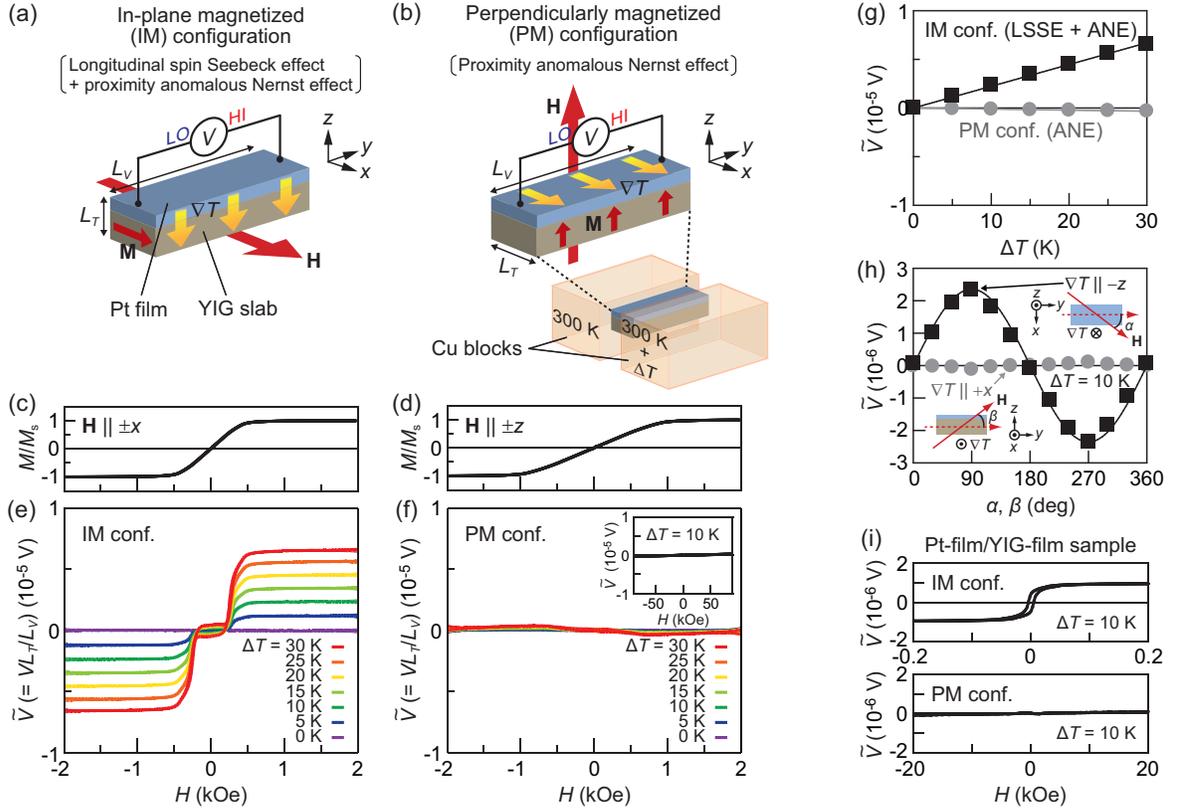}
\caption{(a),(b) Schematic illustrations of the Pt-film/YIG-slab sample in the (a) IM and (b) PM configurations. The sample consists of a single-crystalline YIG slab with the size of $6 \times 2 \times 1$ mm$^3$ and a 10-nm-thick Pt film sputtered on the $6 \times 2$ mm$^2$ surface of the YIG. The temperature gradient was generated by sandwiching the sample between two Cu blocks of which the temperatures were stabilized at $300~\textrm{K}$ and $300~\textrm{K} + \Delta T$. The Pt layer of the sample was electrically well insulated from the Cu blocks by inserting thin insulating (but thermally conductive) sheets between the sample and the Cu blocks. (c),(d) $M$-$H$ curve of the YIG slab, measured when (c) ${\bf H}~|| \pm x$ and (d) ${\bf H}~|| \pm z$. (e),(f) $H$ dependence of $\widetilde{V}$ ($= VL_T/L_V$) in the Pt-film/YIG-slab sample for various values of $\Delta T$ in the (e) IM and (f) PM configurations, where $L_{T} = 1~\textrm{mm}$ ($2~\textrm{mm}$) and $L_{V} = 6~\textrm{mm}$ ($6~\textrm{mm}$) for the IM (PM) configuration. The inset to (f) shows the $H$ dependence of $\widetilde{V}$ at $\Delta T = 10~\textrm{K}$ in the PM configuration, measured when the magnetic field was swept between $\pm 90~\textrm{kOe}$. (g) $\Delta T$ dependence of $\widetilde{V}$ in the Pt-film/YIG-slab sample at $H = 1.2~\textrm{kOe}$ in the IM and PM configurations. (h) The magnetic-field-angle ($\alpha$, $\beta$) dependence of $\widetilde{V}$ in the Pt-film/YIG-slab sample at $\Delta T = 10~\textrm{K}$ and $H = 1.2~\textrm{kOe}$, measured when $\nabla T~|| -z$ ($\nabla T~|| +x$) and ${\bf H}$ was applied in the $x$-$y$ ($y$-$z$) plane at an angle $\alpha$ ($\beta$) to the $y$ direction. (i) $H$ dependence of $\widetilde{V}$ in the Pt-film/YIG-film sample at $\Delta T = 10~\textrm{K}$ in the IM and PM configurations. The Pt-film/YIG-film sample consists of a 10-nm-thick Pt film and a 4.5-$\mu$m-thick single-crystalline YIG film grown on a GGG substrate with the size of $7 \times 6 \times 0.5$ mm$^3$. }\label{fig:5}
\end{center}
\end{figure*}

The LSSE in the Pt/YIG structure can be separated from the ANE by comparing voltage in an in-plane magnetized (IM) configuration (the conventional LSSE setup) and a perpendicularly magnetized (PM) configuration in which $\nabla T$ was applied in the Pt-film-plane direction [compare Figs. \ref{fig:5}(a) and \ref{fig:5}(b)]. In the PM configuration, the ANE voltage can appear, while the LSSE voltage should disappear due to the ISHE symmetry, enabling the estimation of the ANE contamination in the Pt/YIG structure [see Eq. (\ref{equ:SSE1}) and note that ${\bf J}_{\rm s}~||~{\bm \sigma}$ in the PM configuration]. To compare voltage signals between the IM and PM configurations, we used a Pt-film/YIG-slab sample that comprises a single-crystalline YIG slab with the size of $6 \times 2 \times 1$ mm$^3$ covered with a 10-nm-thick Pt film, without substrates to avoid a thermal-conductivity-mismatch problem~\cite{SSE11,SSE13}. Then, the sample was sandwiched between two large Cu blocks of which the temperatures were stabilized at $300~\textrm{K}$ and $300~\textrm{K} + \Delta T$ to apply a uniform temperature gradient to the sample [Fig. \ref{fig:5}(b)]. In the IM (PM) configuration, ${\bf H}$ was applied along the $x$ ($z$) direction. As shown in Fig. \ref{fig:5}(d), we found that the magnetic field of $>1~\textrm{kOe}$ is also large enough to align the magnetization of the YIG slab along the ${\bf H}$ direction in the PM configuration, since the YIG sample used here is a thick slab. Note that, in the PM configuration, even a possible tiny temperature gradient perpendicular to the Pt-film plane does not affect the voltage signal, since the Nernst voltage is not generated due to the collinear orientation of the perpendicular temperature gradient and ${\bf H}$. \par
Figures \ref{fig:5}(e) and \ref{fig:5}(f), respectively, show the $H$ dependence of $\widetilde{V}$ in the Pt-film/YIG-slab sample for various values of $\Delta T$ in the IM and PM configurations, measured when ${\bf H}$ was applied perpendicular to both $\nabla T$ and the longest direction of the sample. Here, $\widetilde{V}$ denotes the voltage including geometric factors: $\widetilde{V} = VL_T/L_V$ with $L_{T(V)}$ being the length of the sample in the temperature-gradient (interelectrode) direction [Figs. \ref{fig:5}(a) and \ref{fig:5}(b)] that enables the quantitative comparison of the signals between the IM and PM configurations. We found that the $\widetilde{V}$ signal in the Pt-film/YIG-slab sample in the PM configuration is much smaller than that in the IM configuration [compare Figs. \ref{fig:5}(e) and \ref{fig:5}(f) and see also Figs. \ref{fig:5}(g) and \ref{fig:5}(h)]. The ratio of the voltage magnitude between the IM and PM configurations is estimated to be $|\widetilde{V}_{\rm PM}/\widetilde{V}_{\rm IM}| \sim 0.05$ at each $\Delta T$, where $\widetilde{V}_{\rm IM(PM)}$ denotes $\widetilde{V}$ in the IM (PM) configuration at $H = 1.2~\textrm{kOe}$. This result indicates that the ANE contamination in the Pt/YIG sample is, if any, less than 5 \% of the voltage signal, since $\widetilde{V}_{\rm PM}$ potentially contains both the proximity ANE and leaked SSE voltages due to experimental errors, i.e., minimal inclination of the ${\bf H}$ direction. We also observed similar behavior of $\widetilde{V}$ in a Pt-film/YIG-film sample on a GGG substrate [Fig. \ref{fig:5}(i)]. All the results shown above confirm that the voltage signal observed in the IM configuration is due almost entirely to the ISHE voltage induced by the LSSE. \par
In conclusion, we report experiments for the exclusive detection of the LSSE. First, we observed the LSSE in the Au/YIG and Pt/Cu/YIG structures. Since Au and Cu are far from the Stoner ferromagnetic instability, these observations confirm the existence of the LSSE free from the ANE induced by the magnetic proximity. In the conventional Pt/YIG structure, the voltage measurements in the in-plane and perpendicularly magnetized configurations show that the LSSE provides a dominant contribution and that the ANE contamination is negligibly small. Whether or not proximity ferromagnetism exists in Pt cannot be discussed based on the present results. However, we can conclude that the observed LSSE voltage is at least irrelevant to the proximity ANE. \par
The authors thank Y. Sakuraba and K. Takanashi for their assistance in magnetometry measurements. This work was supported by PRESTO-JST ``Phase Interfaces for Highly Efficient Energy Utilization'', CREST-JST ``Creation of Nanosystems with Novel Functions through Process Integration'', a Grant-in-Aid for Research Activity Start-up (24860003) from MEXT, Japan, a Grant-in-Aid for Scientific Research (A) (24244051) from MEXT, Japan, LC-IMR of Tohoku University, the Murata Science Foundation, the Mazda Foundation, the Sumitomo Foundation, and MOST (No. 2011CB921802). 
\end{document}